\documentclass[]{jfm}

\usepackage{graphicx}
\usepackage{newtxtext}
\usepackage[varvw]{newtxmath}
\usepackage{natbib}
\usepackage{hyperref}
\usepackage{bm}
\usepackage{tikz}
\usepackage{pgfplots}

\newcommand{\aver}[1]{ \left\langle {#1} \right \rangle}

\newcommand{\tens}[1]{ {#1} }

\usepackage{xcolor,hyperref}

\definecolor{lime}{HTML}{A6CE39}
\DeclareRobustCommand{\orcidicon}{
	\begin{tikzpicture}
	\draw[lime, fill=lime] (0,0) 
	circle [radius=0.16] 
	node[white] {{\fontfamily{qag}\selectfont \tiny ID}};
	\draw[white, fill=white] (-0.0625,0.095) 
	circle [radius=0.007];
	\end{tikzpicture}
	\hspace{-2mm}
}
	
\foreach \x in {A, ..., Z}{\expandafter\xdef\csname orcid\x\endcsname{\noexpand\href{https://orcid.org/\csname orcidauthor\x\endcsname}
			{\noexpand\orcidicon}}
}

\title[Energy transfer and scale organisation in dense canopy turbulence]{Energy transfer and scale organisation in dense canopy turbulence}
\author{Riccardo Bertoncello
\aff{1}\aff{2}\corresp{\email{riccardo.bertoncello@polimi.it}},
Alessandro Chiarini
\aff{1}\aff{2}\corresp{\email{alessandro.chiarini@polimi.it}},
Giulio Foggi Rota
\aff{2},
Maurizio Quadrio
\aff{1}
and
Marco Edoardo Rosti
\aff{2}
}
\affiliation{
\aff{1} Dipartimento di Scienze e Tecnologie Aerospaziali, Politecnico di Milano, via La Masa 34, 20156 Milano, Italy
\aff{2} Complex Fluids and Flows Unit, Okinawa Institute of Science and Technology Graduate University, 1919-1 Tancha, Onna-son, Okinawa, 904-0495, Japan
}

\begin{document}
\maketitle

\begin{abstract}

This study identifies the scale-dependent processes that sustain turbulence in dense submerged canopy flows. Using a scale-resolved energy budget, we determine where in space and at which scales production, pressure--strain redistribution, and inter-scale transfer predominantly occur, and how they link the canopy layer to the overlying shear flow.
We show that energy production is localised at the interfacial shear layer, over a narrow range of streamwise and spanwise scales, while fluctuations within the canopy are primarily maintained through inter-scale transfer and pressure--strain redistribution. The dynamically active scales in the canopy are largely imposed by outer-layer structures, with their organisation and coherence mediated by these inherited motions.
Energy exchange across the canopy interface is asymmetric but not unidirectional: although the dominant transfer is from the outer layer towards the canopy, intermittent reverse interactions occur at all scales. The most intense cross-interface exchanges are associated with finer-scale motions rather than large-scale structures, indicating that extreme interfacial energy fluxes are governed predominantly by small-scale dynamics.
The flexibility of the canopy weakens the coherence of outer-layer structures and reduces the efficiency of inter-layer energy transfer, thereby altering both the organisation and scale of the fluctuations within the canopy.
These results clarify how turbulence in dense canopies is organised and sustained, providing a unified energetic interpretation that links coherent structures to scale-dependent mechanisms.

\end{abstract}


\section{Introduction}
\label{sec:introduction}

Canopy flows develop over impermeable surfaces where slender, attached, and often flexible elements form a complex interfacial layer. Such flows are ubiquitous in both natural and built environments, including terrestrial and aquatic vegetation, as well as urban landscapes. In forests, canopy-flow interactions regulate the exchange of carbon dioxide and oxygen \citep{bui-etal-2021}; in aquatic systems, vegetation supports biodiversity and mitigates coastal hazards by damping wave energy \citep{lei-nepf-2019}. In urban contexts, understanding canopy-layer dynamics can inform building designs that improve pollutant dispersion \citep{michioka-takimoto-sato-2014}.

Canopies are typically classified as emergent ($h/H>1$) or submerged ($h/H<1$), depending on the ratio between the element height $h$ and the fluid depth $H$. In emergent canopies, the flow is dominated by filament-wake eddies throughout the entire fluid depth \citep{vanRooijen-etal-2018}. In contrast, submerged canopies exhibit multiscale dynamics, characterised by the coexistence of large-scale motions and smaller-scale turbulence \citep{monti-etal-2022}. For submerged configurations, fluid-structure interactions are strongly influenced by the solidity $\lambda$, i.e., the ratio of frontal to bed area occupied by the canopy \citep{nepf-2012a}. In dense canopies (high $\lambda$), closely spaced elements disrupt the coherence of incoming eddies, limiting their penetration \citep{sharma-garcia-mayoral-2020}, whereas in sparse canopies (low $\lambda$) larger spacings allow overlying turbulent motions to extend down to the wall \citep{sharma-garcia-mayoral-2020a}. Recent findings suggest that solidity alone is insufficient to characterise canopy-induced dynamics, and that additional geometrical parameters must be considered \citep{monti-etal-2022}. In this study, we focus on dense submerged canopies composed of slender filaments ($h \gg d$, with $d$ the cross-sectional stem scale), representative of aquatic vegetation.

Turbulent flow over dense, submerged canopies exhibits pronounced vertical stratification \citep{poggi-katul-albertson-2004}. Analogously to flows over porous media, the domain can be broadly divided into two regions: an inner region within the canopy, where velocity fluctuations are attenuated, and an outer region above the canopy, where the mean velocity profile approaches a logarithmic-like behaviour \citep{chen-garcia-mayoral-2023}. At the canopy tips, a shear layer develops due to the abrupt change in drag \citep{raupach-finnigan-brunei-1996}, which is susceptible to Kelvin--Helmholtz (KH)-like instabilities \citep{finnigan-2000}. The resulting spanwise-coherent KH rolls evolve under mean shear into more complex, three-dimensional hairpin-like vortices, which eventually break down into smaller turbulent structures \citep{bailey-stoll-2016}. These large-scale motions coexist with smaller-scale turbulence and extend their influence into the canopy interior \citep{sharma-garcia-mayoral-2020a, monti-etal-2022, foggirota-etal-2024}. Observations show that the penetration and coherence of these motions are affected by element spacing, stem inclination, and flexibility.

Over the past decades, numerous studies have investigated turbulence in canopy flows using single-point turbulent kinetic energy (TKE) budgets \citep{wilson-shaw-1977, shen-leclerc-1997, lopez-garcia-2001, poggi-katul-albertson-2004, defina-bixio-2005, cui-neary-2008}. These works show that TKE is primarily generated near the canopy tip, where mean shear is strongest, and subsequently redistributed both within the canopy and into the overlying flow. However, single-point measurements cannot capture the scale dependence of these processes or identify the flow structures responsible for production, redistribution, and inter-scale transfer. More recently, \citet{he-liu-shen-2022} applied a spectral TKE budget to relate energy distribution to streamwise Fourier modes. While this provides a first step toward a scale-aware perspective, it remains limited to one-dimensional modes and TKE alone, leaving three-dimensional interactions, anisotropy, and connections to coherent structures unresolved.

Despite evidence of multiscale dynamics, the drivers of turbulence in dense canopies remain unclear. In particular, the scales dominating production, pressure--strain redistribution, and energy exchange across the canopy-tip interface are not yet quantified. A detailed, scale-resolved characterisation connecting energy dynamics to coherent structures is therefore needed.
The objective of this work is indeed to identify the scale-dependent mechanisms that sustain turbulence in dense submerged canopies, and to determine how energy is exchanged between the canopy layer and the overlying flow. Building on the DNS database of \cite{monti-olivieri-rosti-2023}, we analyse the Reynolds stresses budget in the joint space of wall-normal position and separation scale. We employ the anisotropic generalised Kolmogorov equations (AGKE) \citep{gatti-etal-2020a}, which provide scale-space transport equations for each Reynolds stress component and enable a local characterisation of production, pressure--strain redistribution, and inter-scale fluxes. This framework allows us to determine which scales are dynamically active in different regions of the flow, how they mediate canopy-outer coupling, and how these mechanisms relate to coherent structures.
Extending the spectral analysis of \cite{he-liu-shen-2022}, we resolve the full tensorial Reynolds stresses budget and consider both streamwise and spanwise separations, providing a three-dimensional description of scale interactions. Particular attention is devoted to the inner canopy, to identify the scales and structures responsible for sustaining the fluctuations, and to understand the mechanisms governing energy transfer from the shear layer. We also examine how canopy flexibility modulates these structures and their scale organisation.

The remainder of this paper is organised as follows. Section~\ref{sec:methods} outlines the physical setup, the DNS database, and provides a brief overview of the AGKE framework. In Sections~\ref{sec:structure} and \ref{sec:agke-results}, we focus on the case with rigid canopies to identify characteristic flow scales and examine the mechanisms sustaining velocity fluctuations in the combined physical and scale space. The influence of canopy flexibility is then explored in Section~\ref{sec:flex}. Finally, concluding remarks and future perspectives are presented in Section~\ref{sec:conclusions}.

\section{Mathematical formulation}
\label{sec:methods}

\subsection{Problem setup and DNS database}
\label{sec:physical_setup}
\begin{figure}
  \centering
  \includegraphics[width=0.9\textwidth]{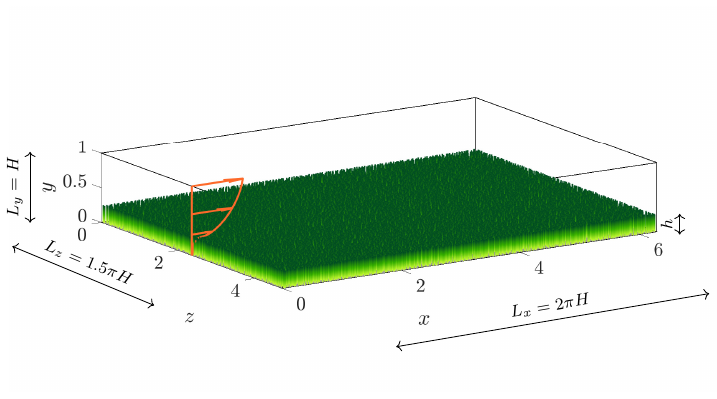}
  \caption{Sketch of the computational domain and of the reference frame.}
  \label{fig:sketch}
\end{figure}
We use the direct numerical simulation (DNS) database of turbulent flow over dense, submerged flexible and rigid canopies generated by \citet{monti-olivieri-rosti-2023}. Full details of the numerical method and computational procedures are provided in \citet{monti-olivieri-rosti-2023} and \citet{foggirota-etal-2024}, where the interested reader can also find validations against experimental measurements.

We study the turbulent flow in a planar half-channel containing a finite number of slender filaments clamped to the bottom wall; see figure~\ref{fig:sketch}. A Cartesian coordinate system is used, with $x$ ($u$), $y$ ($v$), and $z$ ($w$) denoting the streamwise, wall-normal, and spanwise directions (and corresponding velocity components). For simplicity, the alternative notation $x_1$ ($u_1$), $x_2$ ($u_2$), and $x_3$ ($u_3$) is also occasionally adopted.

The fluid motion is governed by the incompressible Navier--Stokes equations for a Newtonian fluid:
\begin{equation}
  \frac{\partial \bm{u}}{\partial t} + \bm{\nabla} \cdot (\bm{u} \bm{u}) = - \frac{1}{\rho_f} \bm{\nabla} p + \nu \nabla^2 \bm{u} + \bm{f}, \quad
  \bm{\nabla} \cdot \bm{u} = 0,
\label{eq:ns}
\end{equation}
where $\bm{u}$ is the velocity vector, $p$ is the pressure, and $\rho_f$ and $\nu$ denote the fluid density and kinematic viscosity, respectively. The term $\bm{f}$ represents the interaction force between the fluid and the filamentous structures.
The bulk Reynolds number is fixed at $Re_b = U_b H / \nu = 5000$ for all cases, where $U_b = (1/\Omega) \int_\Omega u \, \mathrm{d} \Omega$ is the bulk velocity, $\Omega$ is the volume of the computational domain, and $H$ is the channel half-height, measured from the wall to the centerline; see figure~\ref{fig:sketch}. The simulations maintain a constant $U_b$ by suitably adapting in time the pressure gradient.
The flexibility of the filaments is characterised by the dimensionless Cauchy number:
\begin{equation}
    Ca = \frac{1}{2} \frac{\rho_f d h^3 U_b^2}{\gamma},
\end{equation}
where $d$ and $h$ are the filament diameter and length, respectively and $\gamma$ is the bending stiffness. The Cauchy number quantifies the ratio between the fluid-induced deforming force and the elastic restoring force of the filaments, with larger values corresponding to more flexible structures.

The simulations were performed with the solver \href{https://groups.oist.jp/cffu/code}{\textit{Fujin}}, which uses second-order finite differences schemes in space and a second-order Adams-Bashfort scheme in time. A fractional-step method is employed for the temporal advancement of the solution. The computational domain extends for $L_x = 2 \pi H$, $L_y = H$ and $L_z = 1.5 \pi H$. The adequacy of these values are discussed in \cite{monti-olivieri-rosti-2023} and further confirmations are provided by \cite{sathe-giometto-2024}. No-slip and no-penetration boundary conditions are applied at the bottom wall ($y = 0$), symmetry conditions are used at the top boundary ($y = H$), and periodic conditions are enforced in the streamwise and spanwise directions. The computational domain is discretised with $(N_x,N_y,N_z) = (1152,384,864)$ points. In the streamwise and spanwise directions a uniform distribution of points is adopted. In the wall-normal direction, within the canopy the distribution of points is uniform, while above the canopy tip the grid is linearly stretched up to the top boundary.

The canopy is formed by $n_x \times n_z = 144 \times 108$ filaments in the streamwise and spanwise directions. Their length is $h = 0.25H$. 
To avoid preferential flow channeling, the surface of the solid wall is divided in a grid of $n_x \times n_z$ equal squares; each is assigned one filament, whose location within the square is selected randomly. The fluid-solid interaction is dealt with a Lagrangian immersed-boundary method \citep{yu-2005, huang-shin-sung-2007}; see also \cite{marchioli-rosti-verhille-2025}. Each filament is discretised with $N_l$ Lagrangian points uniformly distributed along the stem length, with $N_l=126$ for the rigid filaments ($Ca = 0$) and $N_l=32$ for the flexible ones ($Ca \neq 0$). 
In the rigid case, a greater number of Lagrangian points is employed to resolve the larger velocity difference between the fluid and the structure. This denser distribution results in more uniform forcing, with roughly one Lagrangian point per Eulerian grid point.
The no-slip condition $\partial \bm{X}/\partial t = \bm{u}(\bm{X}(s,t),t)$ is enforced at each Lagrangian point (see \S\ref{appA}). 
An anelastic collision kernel prevents geometrical overlap; however, within the present parameter range, contacts are extremely rare, and disabling the collision model produces statistically indistinguishable results. At every time-step the fluid velocity $\bm{u}$ is interpolated on the Lagrangian points, and the fluid-structure force acting on the filaments is evaluated as
\begin{equation}
    \bm{f}_{l} = - \beta  \left( \bm{u}_{l} - \frac{\partial \bm{X}}{\partial t} \right),
\end{equation}
where $\partial \bm{X}/\partial t$ is the velocity of the filament, $\bm{u}_{l}$ is the fluid velocity interpolated on the position of the filament, and $\beta$ is a properly chosen proportionality coefficient which is set equal to $10$. Note that we use the subscript ``$l$'' for quantities defined in the Lagrangian frame of reference.
The fluid-structure force $\bm{f}_{l}$ is transferred to the Eulerian grid via a spreading function, producing the volume force $\bm{f}$ appearing in equation \eqref{eq:ns}. 
This approach enforces the no-slip and no-penetration conditions over a finite region rather than at isolated points. 
When the Lagrangian spacing is comparable to the Eulerian grid, overlapping kernels ensure continuous and effective enforcement along the filament.

The structural dynamics of the filaments for $Ca \neq 0$ is described with the generalised Euler--Bernoulli beam model:
\begin{equation}
\begin{cases}
\displaystyle
  \Delta \Tilde{\rho} \frac{\partial^2 \bm{X}}{\partial t^2} = \frac{\partial}{\partial s} \left( T \frac{\partial \bm{X}}{\partial s} \right) - \gamma \frac{\partial^4 \bm{X}}{\partial s^4} - \bm{f}_l, \\ \\
\displaystyle
  \frac{\partial \bm{X}}{\partial s} \cdot \frac{\partial \bm{X}}{\partial s} = 1,
\end{cases}
\label{eq:beam_equation}
\end{equation}
where $\bm{X}(s,t)$ is the position of a point along the filament, function of the curvilinear abscissa $s$ and of the time $t$, $\Delta \Tilde{\rho}$ is the linear density difference between the filament and the fluid, defined as $\Delta \Tilde{\rho} = \left( \rho_s - \rho_f \right)\pi d^2/4$ ($\rho_s$ is the density per unit volume of the filaments), and $T$ is the tension enforcing the inextensibility constraint. 
The lower end of the filaments ($s = 0$) is clamped to the bottom wall, while the other end ($s = h$) is free to oscillate, thus leading to the following set of boundary conditions: $\bm{X}|_{s=0} = \bm{X}_0$, $\partial \bm{X}/\partial s|_{s=0} = (0,\phi_0,0)$, $\partial^2\bm{X}/\partial s^2|_{s=h}=\bm{0}$, $\partial^3\bm{X}/\partial s^3|_{s=h}=\bm{0}$ and $T|_{s=h}=0$, where $\phi_0=0$ is the clamp angle of the stems.
Equation \eqref{eq:beam_equation} is solved following \cite{huang-shin-sung-2007}, except for the bending term $\gamma \partial^4 \bm{X}/\partial s^4$ which is treated implicitly as in \cite{banaei-rosti-brandt-2020} to allow for a larger time step.
\begin{figure}
    \centering
    \includegraphics[width=\textwidth]{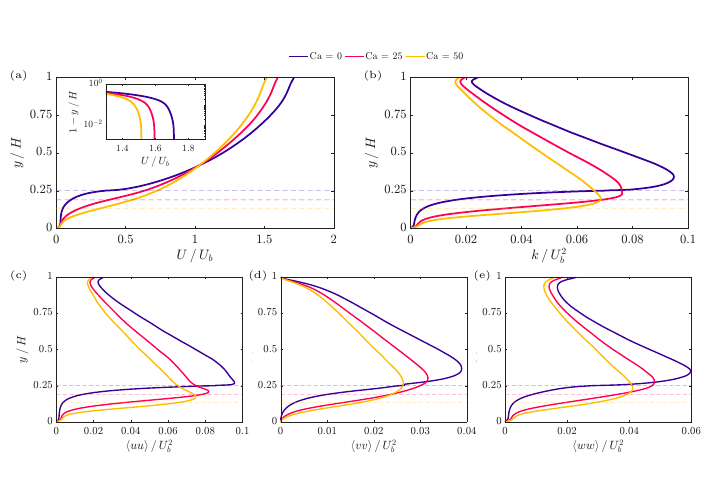}
    \caption{(a) Mean streamwise velocity $U$, (b) turbulent kinetic energy $k$, and (c–e) Reynolds stresses $\aver{uu}$, $\aver{vv}$, and $\aver{ww}$ for $Ca = 0$, 25, and 50. All quantities are non-dimensionalised by $U_b$ and $H$. The dashed lines indicate the average wall-normal location of the tip of the filaments, i.e., $\aver{y_{tip}}_f/H = 0.25, 0.19, 0.135$ for $Ca=0, 25, 50$, respectively; $\aver{\cdot}_f$ denotes average in time and across filaments. The same colour code is used for continuous and dashed lines.}
    \label{fig:first_order_stats}
\end{figure}

In this study, we investigate three values of the Cauchy number, i.e. $Ca = 0$, $Ca = 25$, and $Ca = 50$, representing the rigid, intermediate, and flexible regimes, respectively \citep[see][]{monti-olivieri-rosti-2023}. Figure~\ref{fig:first_order_stats} presents vertical profiles of the mean velocity, turbulent kinetic energy, and Reynolds stresses for these cases.

For comparison, reference data from a planar open-channel flow without a canopy at the same bulk Reynolds number are also included in parts of the analysis (see \S\ref{sec:structure}). The main computational parameters for all simulations are summarised in table~\ref{tab:simulations}.

For all the analysed cases, a total of 50 flow field snapshots, distributed over a time of $30 H/U_b$, are employed for computing the flow statistics. After the Reynolds decomposition, the velocity field is written as the sum of the mean flow $\bm{U}=\aver{\bm{u}}$ and the fluctuations, where the $\aver{\cdot}$ operator represents the statistical average along the homogeneous directions and in time. For simplicity, hereinafter we use capital letters for mean quantities and small letters for fluctuations. 

Unless stated otherwise, all quantities are nondimensionalised using the bulk velocity $U_b$ and the channel half-height $H$. Outer units are preferred for our analysis, as they directly relate to the large-scale motions in the unobstructed region of the channel \citep{peruzzi-etal-2020}. We also use the superscript ``$+$'' to denote normalisation in viscous (inner) units, where quantities are made dimensionless using the friction velocity $u_\tau$ and the kinematic viscosity $\nu$. In canopy flows, when viscous units are applied to quantities in the outer layer, the friction velocity is evaluated at the virtual origin $y_{vo}$, defined as the effective origin of the logarithmic velocity profile. Following \cite{foggirota-etal-2024}, this virtual origin is determined by matching the mean velocity profile above the canopy to a canonical logarithmic law, with the corresponding friction velocity computed as $u_\tau = \left( \nu \mathrm{d}U/\mathrm{d}y - \langle u v \rangle - \int^H_y \aver{f_x}(s) \,\mathrm{d}s \right)^{1/2}\Big|_{y=y_{vo}}$. For the smooth-wall reference case, the friction velocity is instead evaluated directly at the physical wall ($y=0$). Notably, the estimated values of $u_\tau$ and $y_{vo}$ are highly robust; applying alternative identification strategies, such as the diagnostic-function matching proposed by \cite{chen-garcia-mayoral-2023}, yields only marginal differences that do not alter any physical results. Finally, to eliminate any potential coordinate ambiguity, all wall-normal positions throughout this manuscript are consistently expressed as distances from the physical wall ($y$), regardless of whether they are normalised in inner or outer units. For reference, we report the evaluated values of $y^+_{vo}$ for the different configurations in table \ref{tab:simulations}.


\begin{table}
  \begin{center}
\def~{\hphantom{0}}
  \begin{tabular}{lcccccccccc}
    & $Ca$   & $Re_b$  &   $(Lx, Lz)/H$ & $N_x \times N_y \times N_z $ & $n_x \times n_z$ & $\lambda$ & $h/H$ & $\aver{\text{d}p /\text{d}x}$ & $\aver{y_{tip}}_f$ & $y^+_{vo}$ \\[3pt]
    Rigid    & 0  & 5000 & 2$\pi$, 1.5$\pi$ & 1152 $\times$ 384 $\times$ 864 & 144 $\times$ 108 & 1.43 & 0.25 & 0.0441 & 0.25 & 186 \\
    Intermediate & 25 & 5000 & 2$\pi$, 1.5$\pi$ & 1152 $\times$ 384 $\times$ 864 & 144 $\times$ 108 & 1.43 & 0.25 & 0.0277 & 0.19 & 109 \\
    Flexible & 50 & 5000 & 2$\pi$, 1.5$\pi$ & 1152 $\times$ 384 $\times$ 864 & 144 $\times$ 108 & 1.43 & 0.25 & 0.0208 & 0.135 & 71 \\
    Channel & - & 5000 & 2$\pi$, 1.5$\pi$ & 1152 $\times$ 384 $\times$ 864 & - & - & - & 0.0034 & - & - \\
  \end{tabular}
  \caption{Flow and canopy parameters for the rigid and flexible cases. $Re_b$ is the bulk Reynolds number, $L_x$ and $L_z$ are the streamwise and spanwise sizes of the computational domain, $N_x$, $N_y$ and $N_z$ are the numbers of grid points in the streamwise, wall-normal and spanwise directions, $n_x$ and $n_z$ are the numbers of stems in the streamwise and spanwise direction, $\lambda$ is the canopy solidity, $\aver{\text{d}p/\text{d}x}$ is the average streamwise pressure gradient required to sustain the flow, $\aver{y_{tip}}_f$ is the average wall-normal location of the tip of the filaments, with $\aver{\cdot}_f$ denoting average in time and across filaments.}
  \label{tab:simulations}
  \end{center}
\end{table}

\subsection{The anisotropic generalised Kolmogorov equations}
\label{sec:agke_equations}
The anisotropic generalised Kolmogorov equations (AGKE), introduced by \cite{gatti-etal-2020a} and later extended by \cite{gattere-etal-2023}, provide the exact budget equations for the components of the second-order velocity structure function tensor, $\langle \delta u_i \delta u_j \rangle$. Here, $\delta u_i = u_i(\bm{x}_\beta,t) - u_i(\bm{x}_\alpha,t)$ denotes the increment of the fluctuating velocity component $u_i$ between two points $\bm{x}_\alpha$ and $\bm{x}_\beta$.
The midpoint $\bm{X} \equiv (\bm{x}_\alpha + \bm{x}_\beta)/2$ and the separation vector $\bm{r} \equiv \bm{x}_\beta - \bm{x}_\alpha$ define the position and the scale, respectively. In the general case, the structure function tensor $\langle \delta u_i \delta u_j \rangle$ depends on seven independent variables: the three components of $\bm{X}$, the three components of $\bm{r}$, and time $t$.

The tensor $\langle \delta u_i \delta u_j \rangle$ combines single-point and two-point statistics. Specifically, it can be expressed as:
\begin{equation}
\langle \delta u_i \delta u_j \rangle (\bm{X}, \bm{r}, t) = \mathsf{V}_{ij}(\bm{X}, \bm{r}, t) - \mathsf{R}_{ij}(\bm{X}, \bm{r}, t) - \mathsf{R}_{ij}(\bm{X}, -\bm{r}, t),
\end{equation}
where $\mathsf{R}_{ij}(\bm{X}, \bm{r}, t) = \langle u_i(\bm{x}_\alpha,t) u_j(\bm{x}_\beta,t) \rangle$ is the two-point velocity correlation tensor, and $\mathsf{V}_{ij} = \langle u_i u_j \rangle|_{\bm{x}_\alpha} + \langle u_i u_j \rangle|_{\bm{x}_\beta}$ is the sum of the single-point Reynolds stresses at the two $\bm{x}_\alpha$ and $\bm{x}_\beta$ points. The structure function thus reflects the combined effect of local turbulence intensity and the correlation of velocity fluctuations across scales. Notably, $\mathsf{R}_{ij}$ carries the essential scale-dependent information.

A scalar quantity of interest is $\langle \delta q^2 \rangle = \sum_i \langle \delta u_i \delta u_i \rangle$, often interpreted as a proxy for the turbulent kinetic energy associated with eddies of size $r = |\bm{r}|$ at position $\bm{X}$, i.e., the scale energy. However, this interpretation is only approximate: $\langle \delta q^2 \rangle$ reflects the cumulative energy of all eddies with scales up to $r$, rather than the energy at scale $r$ itself. In the idealised case of homogeneous isotropic turbulence, where statistics depend only on $r$, this becomes evident from the identity $\langle \delta q^2 \rangle = 2 \mathsf{V}_{ii} - 2 \mathsf{R}_{ii}(r)$, so that $\langle \delta q^2 \rangle = 0$ at $r=0$ and tends to $2 \mathsf{V}_{ii} $ as $r \to \infty$. However, as discussed by \citet{davidson-pearson-2005}, this behaviour also includes contributions from large-scale eddies due to their associated enstrophy, and thus cannot be regarded as purely local in scale.

The derivation of the AGKE begins with the equation for the velocity difference $\delta u_i$, obtained by subtracting the Navier--Stokes equations at points $\bm{x}_\alpha$ and $\bm{x}_\beta$. The procedure involves multiplication by $\delta u_j$, expressing all terms as two-point differences or sums, transforming the coordinate system from $(\bm{x}_\alpha, \bm{x}_\beta)$ to $(\bm{X}, \bm{r})$, and averaging. Full details are provided in \cite{gatti-etal-2020a} and \cite{gattere-etal-2023}.
The AGKE describe the scale-by-scale balance of Reynolds stresses, linking the temporal evolution of $\langle \delta u_i \delta u_j \rangle$ at a given scale and position to the combined effects of production, transport, and dissipation. In the present study, the flow is statistically stationary and homogeneous in two directions, reducing the number of independent variables to four: $(Y, r_x, r_y, r_z)$. The AGKE tailored to this configuration read

\begin{equation}\label{eq:agke_channel}
\begin{gathered}
    \frac{\partial}{\partial r_k} \biggl( \underbrace{ \langle \delta u_i \delta u_j \delta U_k \rangle }_{\phi^{m}_{k,ij}} + 
    \underbrace{ \langle \delta u_i \delta u_j \delta u_k \rangle }_{\phi^{t}_{k,ij}} -
    \underbrace{2\nu \frac{\partial}{\partial r_k} \langle \delta u_i \delta u_j \rangle}_{\phi^{v}_{k,ij}} \biggr) + \\ 
    \frac{\partial}{\partial Y} \biggl( \underbrace{ \langle v^* \delta u_i \delta u_j \rangle }_{\psi^{t}_{ij}} + 
    \underbrace{ \frac{1}{\rho_f} \langle \delta p \delta u_j \rangle \delta_{i2} + \frac{1}{\rho_f} \langle \delta p \delta u_i \rangle \delta_{j2} }_{\psi^{p}_{ij}} - 
    \underbrace{ \frac{\nu}{2} \frac{\partial}{\partial Y} \langle \delta u_i \delta u_j \rangle }_{\psi^{v}_{ij}} \biggr) = \\
    \underbrace{- \langle v^* \delta u_j \rangle \delta \left( \frac{\text{d}U}{\text{d}y} \right) \delta_{i1} - 
    \langle v^* \delta u_i \rangle \delta \left( \frac{\text{d}U}{\text{d}y} \right) \delta_{j1} - 
    \langle \delta v \delta u_j \rangle \left( \frac{\text{d}U}{\text{d}y} \right)^* \delta_{i1} - 
    \langle \delta v \delta u_i \rangle \left( \frac{\text{d}U}{\text{d}y} \right)^* \delta_{j1}}_{\tens{P}_{ij}} + \\
    \underbrace{ \frac{1}{\rho_f} \biggl \langle \delta p \frac{\partial \delta u_i}{\delta X_j} \biggr \rangle + \frac{1}{\rho_f} \biggl \langle \delta p \frac{\partial \delta u_j}{\delta X_i} \biggr \rangle }_{\Pi_{ij}} -    
    \underbrace{4\epsilon^*_{ij}}_{\tens{D}_{ij}} +
    \underbrace{ \langle \delta u_j \delta f_i \rangle + \langle \delta u_i \delta f_j \rangle }_{\tens{F}_{ij}},
\end{gathered}    
\end{equation}
where $i = 1,2,3$ and $j=1,2,3$. Here $\delta_{ij}$ is the Kronecker delta, the superscript ``$*$'' denotes the average of a given quantity between the two points $\bm{x}_\alpha$ and $\bm{x}_\beta$, and $\epsilon_{ij} = \nu \aver{\partial u_i/\partial x_k \ \partial u_j/\partial x_k}$ is the pseudo-dissipation tensor. The $\bm{\phi}^{m}_{ij}$, $\bm{\phi}^{t}_{ij}$ and $\bm{\phi}^{v}_{ij}$ terms are the mean, turbulent and viscous fluxes in the space of scales.
The ${\psi}^{t}_{ij}$, ${\psi}^{p}_{ij}$ and ${\psi}^{v}_{ij}$ terms, instead, are the turbulent, pressure and viscous fluxes in the physical space; they describe how energy is transported in the wall-normal direction.
In addition, $\tens{P}_{ij}$, $\Pi_{ij}$, $\tens{D}_{ij}$ and $\tens{F}_{ij}$ are the production, pressure-strain, dissipation and forcing terms, respectively. All the listed terms are named in analogy to the single-point budgets, see for example \cite{pope-2000}. The production term $\tens{P}_{ij}$ couples the mean and fluctuating fields and describes the scales and positions at which they exchange energy; the pressure-strain term $\Pi_{ij}$ couples pressure and velocity derivatives, and describes the energy redistribution among the three normal stresses; the forcing term $\tens{F}_{ij}$ describes the effect of the interaction between fluid and filaments in the combined space of scales and positions.
Summing together the terms on the right-hand side yields the source term $\xi_{ij} = \tens{P}_{ij} + \Pi_{ij} + \tens{D}_{ij} + \tens{F}_{ij}$. The sum of the two-points budgets for the three normal stresses leads to the generalised Kolmogorov equation \citep{hill-2002a}. 

In this work, the terms of the AGKE are computed using a computer code, written in the CPL language \citep{luchini-2021}, and originally developed for the isotropic version of the AGKE by \cite{gatti-etal-2019}. For maximum computational efficiency, whenever possible, the code computes two-points correlations in the spectral domain, leveraging Parseval's Theorem.

\section{Results}

We investigate the mechanisms sustaining turbulence in dense canopy flows by analysing the dynamics both in physical space and across scales.
We first identify in \S\ref{sec:structure} the dominant flow scales and their associated coherent structures. Building on this framework, \S\ref{sec:agke-results} examines the production, pressure--strain redistribution, and inter-scale transfer of velocity fluctuations, assessing how these mechanisms vary in space and how they relate to the underlying flow structures.
The rigid canopy configuration ($Ca = 0$) is adopted as the reference case, while the effect of filament flexibility is discussed separately in \S\ref{sec:flex}. For comparison, data from a smooth open-channel flow at matched $Re_b$ are included in \S\ref{sec:structure} to isolate and highlight the impact of the canopy on the dominant scales and coherent motions.

\subsection{Flow structure}
\label{sec:structure}

In this section, we identify the dominant scales in the flow and relate them to coherent structures. Structure functions and two-point correlations are first employed to quantify the characteristic velocity scales. Conditional averaging and filtering are then used to associate these scales with the underlying motions, including KH-like rollers and hairpin vortices. This approach combines objective, scale-resolved diagnostics with physically interpretable representations of the coherent dynamics.

\subsubsection{Structure functions and flow scales}
\label{sec:strfun}

\begin{figure}
  \centering
  \includegraphics[width=0.9\textwidth,trim={0cm 0cm 0cm 0cm},clip]{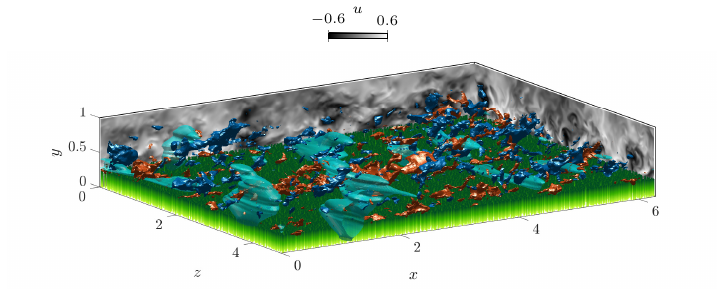}
  \caption{Coherent structures in a snapshot of the canopy flow ($Ca=0$). Isosurfaces of streamwise velocity fluctuations at $u = \pm 0.6$ are coloured in red and blue; isosurfaces of negative filtered pressure fluctuations $\tilde{p} = -0.06$ (see \S\ref{sec:eductions}) are in cyan. The background planes plot streamwise velocity fluctuations $u$ in grayscale.}
  \label{fig:all-structures-new}
\end{figure}

The canopy flow exhibits a broad range of scales, with turbulence largely confined to the outer layer and the interior remaining relatively quiescent (figure \ref{fig:all-structures-new}). Peak turbulent activity occurs near the canopy tips, where fluid-structure interactions generate intense shear layers. Coherent structures observed include low- and high-speed streaks of streamwise velocity (LSS/HSS), along with large-scale vortical motions aligned both in the streamwise and spanwise directions \citep[see also][]{finnigan-2000}.

\begin{figure}
  \centering
  \includegraphics[width=\textwidth,trim={0cm 0cm 0cm 0cm},clip]{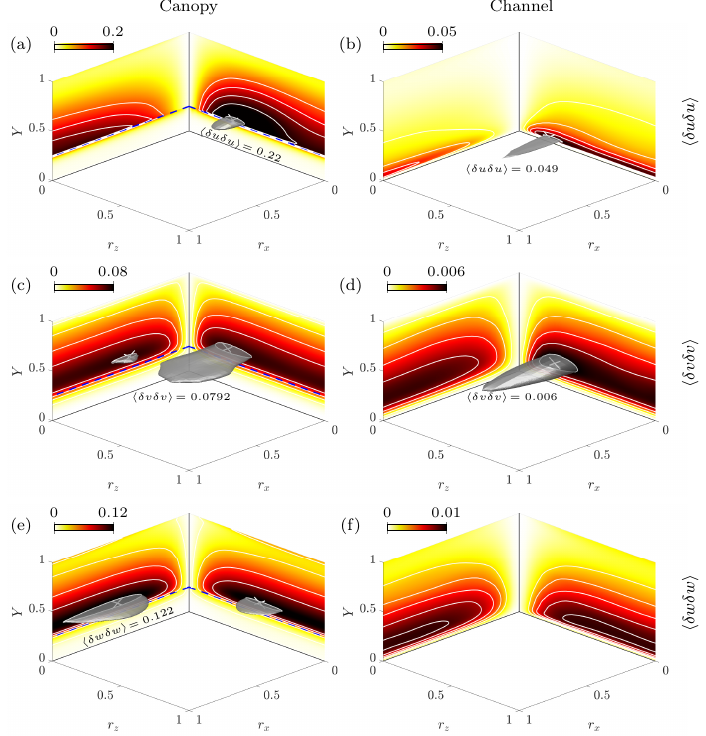} 
  \caption{Structure functions $\langle \delta u_i \delta u_j \rangle$ evaluated in the $r_y = 0$ space. Left column: canopy flow ($Ca=0$); right column: smooth open-channel flow. From top to bottom, rows correspond to $\langle \delta u \delta u \rangle$, $\langle \delta v \delta v \rangle$, and $\langle \delta w \delta w \rangle$, respectively. Grey isosurfaces highlight regions of intense structure function magnitude, corresponding to local peaks of $\langle \delta u_i \delta u_j \rangle$ observed in the $r_y = r_z = 0$ and $r_x = r_y = 0$ planes, and associated with negative values of $\mathsf{R}_{ij}$ at those scales. The values are reported in the figures. The absence of emerging isosurfaces from these planes indicates that no pronounced local peaks of $\langle \delta u_i \delta u_j \rangle$ and $\mathsf{R}_{ij}$ are present at the corresponding scales.}  
  \label{fig:dui-duj-RY0} 
\end{figure}

Figure~\ref{fig:dui-duj-RY0} quantifies the dominant scales through the diagonal components of the structure-function tensor $\langle \delta u_i \delta u_j \rangle$ in the $r_y = 0$ plane, with smooth-wall open-channel data at matched $Re_b$ provided for reference. Peaks in these distributions identify characteristic length scales and highlight the wall-normal variation of the fluctuations. 

Turbulent activity in the outer layer peaks near the canopy tips, where the diagonal components of the structure function tensor reach their maximum, and gradually decays toward the channel centre. At this Reynolds number, turbulence in the core is weak, and remains strongly attenuated within the canopy. Comparison with smooth-wall channel flow reveals both similarities and differences: the streamwise and wall-normal structure functions exhibit distinct peaks at $(Y, r_x, r_z) = (0.273, 0, 0.327)$ and $(0.365, 0, 0.284)$ for the canopy, and at $(0.049, 0, 0.218)$ and $(0.209, 0, 0.262)$ for the smooth-wall channel, corresponding in viscous units to $(Y^+, r_x^+, r_z^+) = (257, 0, 308)$ and $(344, 0, 268)$ for the canopy, and $(14, 0, 64)$ and $(61, 0, 77)$ for the smooth wall. 
These values indicate coherent low- and high-speed streaks (LSS/HSS) and streamwise-aligned vortices in both flows, consistent with observations in high-$Re$ flexible canopies \citep{lohrer-frohlich-2025}. Peaks in the structure functions correspond to minima in the two-point correlation functions $\mathsf{R}_{ij}$, with negative $\mathsf{R}_{22}$ at $r_z \ne 0$ reflecting upwash and downwash motions induced by streamwise-aligned vortices. In canopy flows, these motions are associated with the legs of hairpin vortices, which populate the region just above the tip of the filaments; see \cite{finnigan-shaw-patton-2009} and \S\ref{sec:eductions}.
The coherence of these fluctuations is quantified via correlation coefficients $\rho_{ij}(Y,r_x,r_z) = \mathsf{R}_{ij}(Y,r_x,r_z)/\aver{ u_i u_j }(Y)$ at the structure-function peaks, yielding $\rho_{11}=-0.168$ and $\rho_{22}=-0.04$ for the canopy, and $\rho_{11}=-0.150$ and $\rho_{22}=-0.131$ for the smooth wall. This suggests that LSS/HSS are more coherent in canopy flows than in canonical channel flows, while the opposite is true for streamwise-aligned vortices. 

A clear distinction between the outer-layer canopy flow and the smooth-wall channel is evident in the $r_z = 0$ plane (figure~\ref{fig:dui-duj-RY0}c,d). In the canopy case, the map of $\langle \delta v \delta v \rangle$ displays a secondary, weaker peak at $Y \approx 0.365$ ($Y^+ \approx 344$) and streamwise separation $r_x = 0.436$ ($r^+_x = 411$), which is absent in the smooth-wall configuration. This feature corresponds to spanwise-aligned KH-type vortices originating from the shear-layer instability near the canopy tip (see \S\ref{sec:eductions}), which induce vertical motions along their sides, producing negative $\mathsf{R}_{22}$ for $r_x \ne 0$. The associated correlation coefficient, $\rho_{22}=-0.024$, indicates that these KH structures are less coherent and statistically weaker than the primary LSS/HSS streaks and streamwise-aligned vortices.  

\begin{figure}
  \centering
  \includegraphics[width=\textwidth]{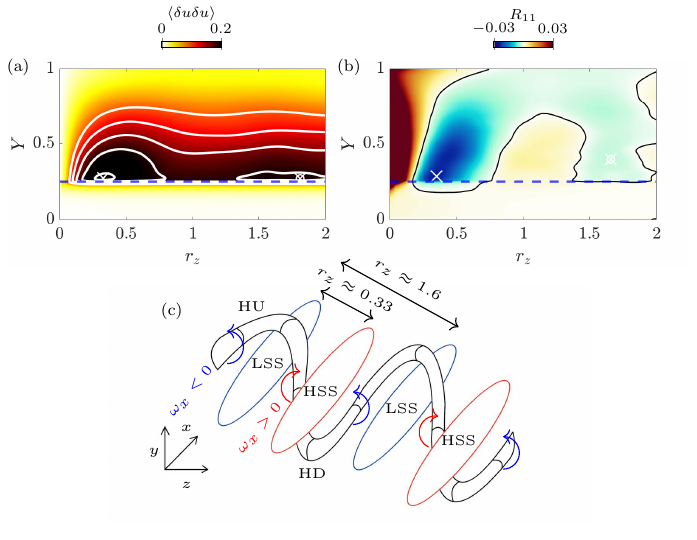} 
  \caption{Top: Contours of $\aver{\delta u \delta u}$ (a) and $\mathsf{R}_{11}$ (b) in the $(Y,r_z)$ space for $Ca=0$. The white crosses indicate the location of the maxima. The white crosses with a circle indicate the location of the secondary maxima. In (b) the black solid line represents the $\mathsf{R}_{11}=0$ contour. The blue dashed line is the wall-normal position of the canopy tip. (c) Schematic representation of the process which leads to the formation of two couples of streaks.}
  \label{fig:dudu-sec-peak}
\end{figure}
A closer inspection of $\aver{\delta u \delta u}$ in the $r_x = r_y = 0$ plane (figure~\ref{fig:dudu-sec-peak}) reveals a secondary peak at larger spanwise separations ($r_z \approx 1.6$, $r_z^+ \approx 1508$) in the canopy, absent in the smooth-wall case. At $Y = 0.28$ ($Y^+ = 264$), the corresponding correlation coefficient is $\rho_{11} = -0.018$. This secondary peak reflects paired high- and low-speed streaks (HSS-LSS), resulting from a spanwise modulation of KH rolls induced by the mean shear near the canopy tips, as illustrated schematically in figure~\ref{fig:dudu-sec-peak}(c) and discussed in \cite{finnigan-2000,finnigan-shaw-patton-2009,bailey-stoll-2016,tschisgale-etal-2021}.  
Peaks in $\aver{\delta w \delta w}$ in the $r_x = 0$ plane further confirm the lateral organisation associated with these KH-induced streaks. This mechanism contrasts with smooth-wall turbulence, where streaks are primarily driven by vertical motions from quasi-streamwise vortices \citep{jimenez-2018}, and no secondary peak in $\aver{\delta u \delta u}$ or $\mathsf{R}_{11}$ is observed.  

\begin{figure}
  \centering
  \includegraphics[width=\textwidth,trim={0cm 0.5cm 0cm 0cm},clip]{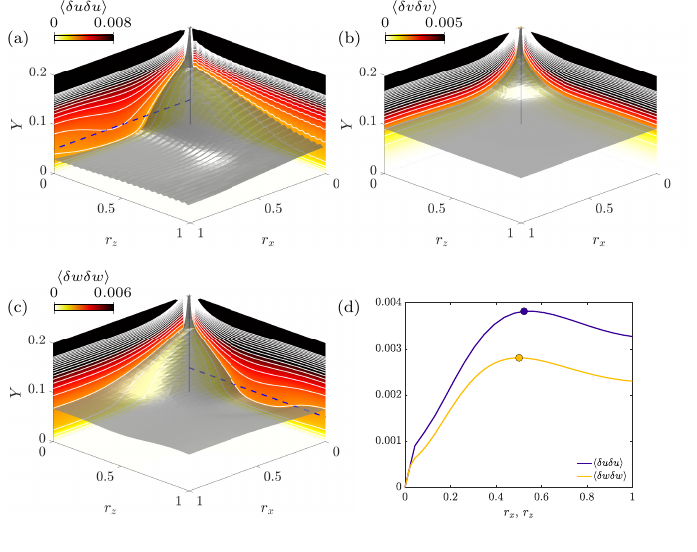}
  \caption{Organisation of velocity fluctuations within the canopy layer ($Ca=0$). Panels (a–c): Structure functions $\aver{\delta u_i \delta u_j}$ shown in the $r_y=0$ space, zoomed inside the canopy region $Y<0.2$. Panel (d): Profiles of $\aver{ \delta u_i \delta u_j }$ along the blue dashed lines indicated in panels (a–c).}
  \label{fig:dui-duj-RY0-zoom}
\end{figure}
Within the inner canopy layer ($Y \le 0.2$; figure~\ref{fig:dui-duj-RY0-zoom}), velocity fluctuations are dominated by the streamwise and spanwise components, consistent with a quasi-two-dimensional flow imposed by the impermeable wall. As expected, $u$-fluctuations are stronger than $w$-fluctuations. Despite overall turbulence suppression, weak but discernible peaks remain in the wall-parallel structure functions, reflecting residual scale organisation: $\aver{\delta u \delta u}$ peaks at $r_x \approx 0.4$--$0.6$, $r_z = 0$, while $\aver{\delta w \delta w}$ peaks at $r_x = 0$, $r_z \approx 0.3$--$0.5$. The corresponding correlation coefficients, $\rho_{11} = \rho_{33} = -0.2$, indicate that these structures are statistically significant and dominate the local fluctuations relative to the quiescent background.
This organisation reflects the modulation imposed by scale-dependent motions above the canopy, as detailed in \S\ref{sec:agke-results}. Streamwise- and spanwise-aligned vortices generate intense sweep events that induce localised downwash into the canopy. These vertical motions are redirected into wall-parallel components via pressure-strain interactions at the wall: sweeps associated with streamwise-aligned vortices enhance spanwise fluctuations, while those associated with spanwise-aligned vortices enhance streamwise fluctuations, reproducing the characteristic scales observed in $\aver{\delta w \delta w}$ and $\aver{\delta u \delta u}$, respectively.

\begin{figure}
  \centering
  \includegraphics[width=\textwidth]{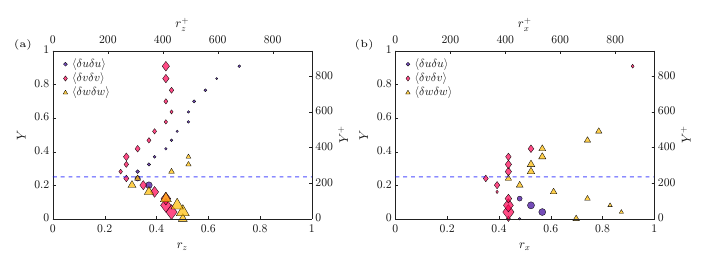}
  \caption{Streamwise and spanwise scales at which $\langle \delta u_i \, \delta u_j \rangle$ attain their maximum values as a function of the wall-normal location $Y$ for $Ca=0$. The dashed blue line indicates the position of the canopy tip. The symbol size corresponds to the correlation coefficient, with larger (smaller) symbols indicating higher (lower) correlation values.}
  \label{fig:scales-strfun}
\end{figure}  
Figure~\ref{fig:scales-strfun} summarises the characteristic length scales as a function of $Y$, along with the corresponding correlation coefficients. Near the canopy tips, structures are compact; moving into the canopy or toward the outer layer, scales increase, reflecting a broadening of influence with distance. Correlation coefficients peak near the wall, demonstrating that the flow structures above the canopy imprint a coherent organisation on the otherwise quiescent interior.

\subsubsection{Eduction of the flow structures}
\label{sec:eductions}

In this section, we complement the statistically robust analysis of structure functions with the identification of the coherent flow structures. 
The goal is to relate the characteristic scales identified previously to specific flow features, providing a qualitative interpretation of their spatial organisation. 
While the previous section delivered quantitative information, the present analysis uses conditional averaging and filtering to isolate representative structures, yielding primarily qualitative insight.

\begin{figure}
\centering
\includegraphics[width=\textwidth,trim={0cm 0.1cm 0cm 0cm},clip]{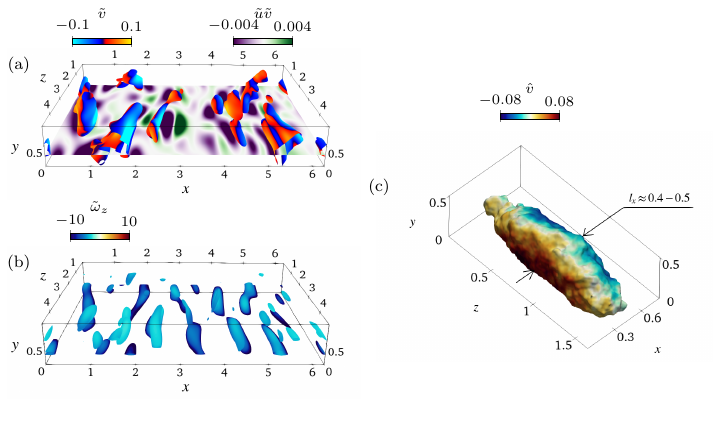}
\caption{Instantaneous and conditionally averaged KH rollers for $Ca = 0$.  
(a) Iso-surfaces of pressure fluctuations $\tilde{p} = -0.04$, coloured by the wall-normal velocity $\tilde{v}$. The horizontal slice shows contours of $\tilde{u}\tilde{v} < 0$ at $y \approx h$.  
(b) Contours of the $Q$-criterion ($\tilde{Q} = 1.5$) computed from the complete (mean plus fluctuating) Fourier-filtered field, coloured by the spanwise vorticity $\tilde{\omega}_z$. In both panels, the mean flow is directed from left to right.  
(c) Conditionally averaged KH roller. The contour of negative pressure $\hat{p} = -0.065$ is shown, coloured by the wall-normal velocity $\hat{v}$.
}
\label{fig:spanwise-rolls}
\end{figure}
KH-like rolls are extracted from the instantaneous flow field by filtering out small-scale fluctuations using a Fourier filter, with filtered quantities denoted by $\tilde{\cdot}$. A Fourier transform is applied in the homogeneous streamwise and spanwise directions, retaining modes with wavenumbers $\kappa_x \le \kappa_{x,\mathrm{th}} = 12$ and $\kappa_z \le \kappa_{z,\mathrm{th}} = 3$, chosen to capture the characteristic streamwise spacing and large-scale spanwise modulation of the rolls \citep{raupach-finnigan-brunei-1996, bailey-stoll-2016}. Moderate variations in the cutoffs do not qualitatively alter the results. The filtered field retains approximately $5\%$ of the wall-normal kinetic energy, providing a measure of the rolls’ energetic footprint.
Figure~\ref{fig:spanwise-rolls} illustrates the filtered flow. Iso-surfaces of negative pressure $\tilde{p}$ (panel a) and positive $Q$ (panel b) identify the KH roll cores, confirming spanwise-coherent structures above the canopy. Contours of $\tilde{u}$ and $\tilde{v}$ show a clear correlation between large-scale streamwise and vertical motions, linking the rolls to sweep and ejection events at the canopy tips and facilitating momentum exchange across the canopy interface.

A qualitative characterisation is obtained via conditional averaging. Events are defined as local minima in the filtered pressure, $\tilde{p} \le \tilde{p}_{\mathrm{th}} = -0.065$, within $y \in (0.25, 0.4)$; conditionally averaged quantities are denoted by $\hat{\cdot}$. Small variations in $\tilde{p}_{\mathrm{th}}$ do not qualitatively affect the results. The resulting KH roll (figure~\ref{fig:spanwise-rolls}c) accounts for roughly $3\%$ of the wall-normal kinetic energy. The negative $\hat{p}$ contour confirms spanwise coherence, while the conditional velocity field exhibits upward motion at the rear and downward motion at the front, consistent with negative spanwise vorticity. Despite the dependence on the chosen $\hat{p}$ contour, the structure shows a streamwise length $\ell_x \approx 0.4-0.5$ ($\ell_x^+ \approx 377-471$), in agreement with the scale inferred from the $\aver{\delta v \delta v}$ map at $r_y = r_z = 0$.

\begin{figure}
\centering
\includegraphics[width=\textwidth,trim={0cm 0cm 0cm 0cm},clip]{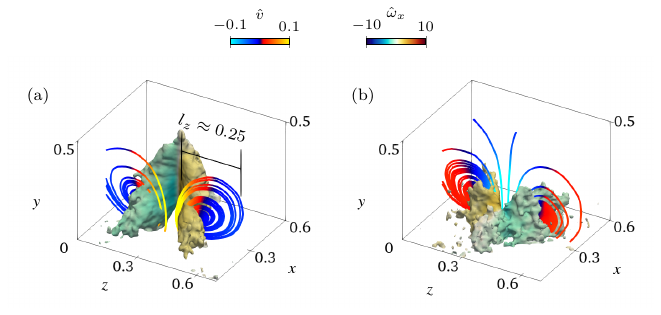}
\caption{The educed hairpin-like vortices for $Ca=0$. Panel (a) shows head-up hairpin-like eddies, panel (b) shows head-down ones. The iso-surface corresponds to $\hat{Q} = 1$, coloured with streamwise vorticity $\hat{\omega}_x$. The streamlines of conditionally averaged two-dimensional velocity, i.e. $\boldsymbol{\hat{u}}_{2D} = (\hat{v}, \hat{w})$ are colored with the wall-normal velocity component $\hat{v}$.}
\label{fig:hairpins}
\end{figure}
Hairpin-like vortices often emerge from the breakdown of KH rolls under mean shear \citep{finnigan-shaw-patton-2009}, although they can also form via longitudinal stretching of weak vorticity between adjacent KH rolls \citep{corcos-lin-1984, neu-1984}. The resulting vortices feature streamwise-aligned, counter-rotating legs.
We identify such events using a conditional averaging strategy analogous to that for KH rolls, applied to the unfiltered velocity field at wall-normal distances $y \in (0.25, 0.4)$. Hairpin-like eddies are detected as local minima of the shear stress, $uv < uv_{th}$ \citep{bailey-stoll-2016}, with the threshold set to $uv_{th}/u^{*2} = -12$, where $u^{*2} = - \min_{y \in [0,H]} \aver{uv}$. Variations in the threshold within $uv_{th}/u^{*2} \in [-14, -10]$ produce qualitatively similar results.
This method captures vortices that transport low- or high-momentum fluid between their legs depending on rotation, generating strong negative $uv$ events \citep{brunet-2020, shig-etal-2023}. Vortices are classified as head-up ($v > 0$, ejections) or head-down ($v < 0$, sweeps) based on the sign of $v$ at the $uv$ minimum.
Figure~\ref{fig:hairpins} shows the resulting conditionally averaged hairpin-like structures. The geometry clearly displays the characteristic counter-rotating legs. The characteristic spanwise scale of their legs is $\ell_z \approx 0.25$ ($\ell_z^+ \approx 236$), consistent with the dominant scale inferred from the $\aver{\delta v \delta v}$ map at $r_x = r_y = 0$. These conditional fields contain approximately $2.5\%$ of the vertical kinetic energy. Some sensitivity of $\ell_z$ to the selected $Q$-criterion contour level is noted.

To place the energetic footprint of the KH rolls and hairpin-like structures in context, we compare them with the quasi-streamwise vortices (QSVs) of a canonical turbulent plane channel flow at $Re_\tau = 180$, isolated using the eduction procedure of \citet{gallorini-quadrio-gatti-2022}. The conditionally averaged QSVs account for approximately $7\%$ of the total turbulent kinetic energy and $5.5\%$ of its wall-normal component---values directly comparable to the global percentages obtained for the KH rolls and hairpin-like structures in the present canopy flow.
These seemingly modest global fractions do not imply limited dynamical significance; indeed, two-point correlations and scale-by-scale energy analyses confirm that the identified structures in both flows actively mediate vertical momentum transport and inter-scale energy transfer. 
Furthermore, because wall-bounded turbulence is strongly inhomogeneous in the wall-normal direction, global energy normalisation inherently masks the intensity of localised coherent motions. When the energetic contribution of the KH rolls is re-evaluated locally---using spatial integrals restricted to horizontal planes or thin three-dimensional slabs ($\Delta y \le 0.1H$) centered around the vortex cores---their contribution to the local wall-normal kinetic energy increases to approximately $8\%$.

\subsection{Sustaining mechanisms}
\label{sec:agke-results}

Having identified the dominant scales, we now examine the production, redistribution and transfer of turbulent kinetic energy towards dissipative scales, together with the energetic coupling between the outer flow and the canopy. We also analyse the mechanisms shaping the structure of the velocity fluctuations within the canopy.

To facilitate the analysis, we focus on the reduced space at $r_y = 0$, which retains the essential physics while simplifying the problem. In this space, the governing equations preserve their conservative form:
\begin{equation}
  \frac{\partial \phi_{x,ij}}{\partial r_x} +
  \frac{\partial \phi_{z,ij}}{\partial r_z} +
  \frac{\partial \psi_{ij}}{\partial Y} = 
  \zeta_{ij}(r_x, r_z, Y),
\end{equation}
demonstrating that in the $r_y=0$ space the evolution of the fluxes is driven by the modified source term:
\begin{equation}
\begin{aligned}
  \zeta_{ij} = &
  - \langle \delta v \delta u_j \rangle \left( \frac{\mathrm{d}U}{\mathrm{d}y} \right) \delta_{i1}
  - \langle \delta v \delta u_i \rangle \left( \frac{\mathrm{d}U}{\mathrm{d}y} \right) \delta_{j1}
  + \frac{1}{\rho_f} \langle \delta p \frac{\partial \delta u_i}{\partial X_j} \rangle
  + \frac{1}{\rho_f} \langle \delta p \frac{\partial \delta u_j}{\partial X_i} \rangle \\
  & + \langle \delta u_i \delta f_j \rangle
  + \langle \delta u_j \delta f_i \rangle
  - 4 \epsilon_{ij}
  - \frac{\partial \phi_{y,ij}}{\partial r_y}.
\end{aligned}
\end{equation}

We analyse the scale energy budget, $\langle \delta q^2 \rangle = \sum_i \langle \delta u_i \delta u_i \rangle$, to identify the scales and locations of energy production and dissipation, and to highlight inter-scale transfers. The source term, $\xi = \sum_i \xi_{ii}$, captures the scale-dependent balance between production and dissipation, while the flux terms quantify transfers across scales and physical space. Near-wall pressure-strain redistribution and interfacial energy transfer are analysed to elucidate the mechanisms underlying the coherent motions observed within the canopy layer (\S\ref{sec:structure}).

\subsubsection{Sources and sinks}
\label{sec:fluxes}

\begin{figure}
  \centering
  \includegraphics[width=0.9\textwidth]{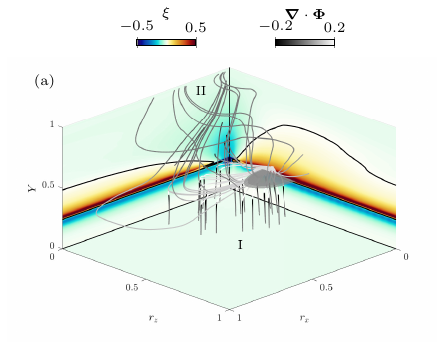} \\
  \vspace{-25pt}
  \includegraphics[width=0.8\textwidth]{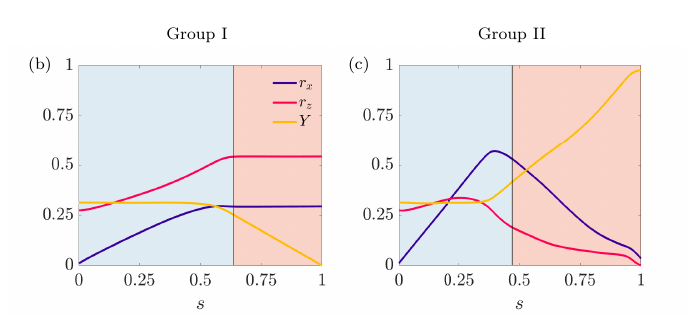} \\
  \vspace{-18pt}
  \includegraphics[width=\textwidth,trim={0cm 0cm 0cm 0cm},clip]{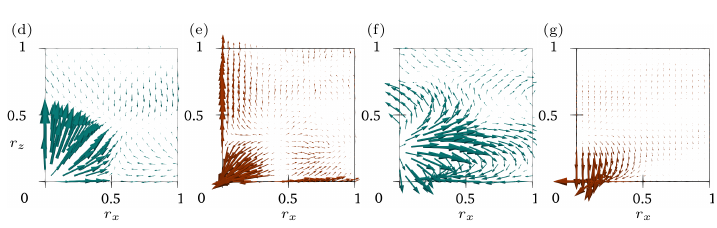}  
  \caption{Panel (a): colour map of the source term $\xi$ in the $r_y = 0$ plane for $Ca = 0$. Field lines of the flux vector $\boldsymbol{\Phi} = (\phi_x, \phi_z, \psi)$ are shown in the same plane, coloured by the divergence $\boldsymbol{\nabla} \cdot \boldsymbol{\Phi}$. The grey surface represents the contour $\xi = 0.9\xi_{\max}$. Panels (b-c): evolution of the values of $r_x$, $r_z$ and $Y$ along a representative field line of $\bm{\Phi}$ for group (a) I and (b) II. The dimensionless arch length $s$ is defined as $s=1/s_{max} \int^{s_{max}}_{0} \text{d}s$, with $\text{d}s = \sqrt{\text{d}r^2_x + \text{d}r^2_z + \text{d}Y^2}$. The red background is for $\bm{\nabla} \cdot \bm{\Phi} < 0$, and the blue one for $\bm{\nabla} \cdot \bm{\Phi} > 0$. Panels (d-g): vectors of (d, f) turbulent transport $\boldsymbol{\Phi}^{t}$ and (e, g) viscous diffusion $\boldsymbol{\Phi}^{v}$ in the $r_y = 0$ plane, shown at different wall-normal locations for the rigid canopy. Panels (d) and (e) correspond to $Y = 0.15H$ (within the canopy), while panels (f) and (g) correspond to $Y = 0.28H$ (just above the canopy tip).
}
  \label{fig:xi-ry0}
\end{figure}

Figure~\ref{fig:xi-ry0}a presents the scale-energy source term $\xi$ along with field lines of the flux vector $\boldsymbol{\Phi} = (\phi_x, \phi_z, \psi)$. Repeated indices are omitted here and in the remainder of this section for brevity. The sign of $\xi$ identifies regions of energy production ($\xi > 0$) and dissipation ($\xi < 0$). The flux vector describes the redistribution of energy in scale space ($r_x$, $r_z$) and the wall-normal direction ($Y$); its field lines indicate the average direction of energy transfer, while its divergence, $\nabla \cdot \boldsymbol{\Phi}$, quantifies the local net flux. A positive divergence implies that fluxes are locally energised, while a negative divergence indicates that fluxes are compensating for local dissipation.
Field lines of $\boldsymbol{\Phi}$ typically originate in source regions and terminate in sinks, tracing the pathways of energy transfers. However, they do not represent causal trajectories, but rather offer a visual representation of the average energy transfer directions.
 
The source term $\xi$ attains its maximum at small scales within the shear layer immediately above the canopy tips, around $r_x \approx 0$, $r_z \in [0.2, 0.25]$ ($r^{+}_z \in [189, 236]$), and $Y \in [0.25, 0.28]$ ($Y^{+} \in [236, 264]$). This region coincides with the peak of the production term $P_{11}$ (not shown), which, owing to flow symmetries, is the sole contributor to scale-energy production. Here, energy is extracted from the mean shear into streamwise velocity fluctuations, while the cross-stream components are sustained through pressure–strain redistribution.
This energetic hotspot defines the driving range of scales \citep{casciola-etal-2003,cimarelli-deangelis-casciola-2013}, where strong mean shear ($\mathrm{d}U/\mathrm{d}y$) and anisotropy ($\aver{ \delta u \delta v }$) inject energy into the turbulent field. These production scales coincide with the energy-containing scales identified above the canopy in the previous section, confirming their dynamical relevance.
Energy sinks ($\xi < 0$) are predominantly located within the canopy and at the smallest scales. In contrast to planar channel flows \citep{gatti-etal-2020a}, where the minimum of $\xi$ occurs at the wall, the strongest dissipation here is found just below the canopy tips. This shift reflects the intense shear and enhanced velocity gradients at the canopy-flow interface, which displace the primary dissipative activity away from the solid boundary. The absence of positive $\xi$ within the canopy layer indicates that no energy production occurs there; rather, as shown below, the fluctuation structure inside the canopy is largely inherited from motions generated above.

\subsubsection{Scale-space energy transfers}

We examine the transfer of energy from the production region ($\xi>0$) to dissipative regions ($\xi<0$) using the scale-space flux vector $\boldsymbol{\Phi}$. The flux organisation reveals a structured redistribution: energy extracted at finite scales within the shear layer above the canopy is conveyed downward into the canopy and upward toward the smallest scales of the outer flow, qualitatively consistent with planar channel turbulence \citep{cimarelli-deangelis-casciola-2013}.

All flux lines originate from a single topological singularity near $(r_x, r_z, Y) \approx (0, 0.28, 0.3)$, i.e., $(r^{+}_x, r^{+}_z, Y^{+}) \approx (0, 264, 283)$, coinciding with the previously identified driving range. From this site, two distinct pathways emerge: one directs energy downward into the canopy and ultimately to the wall (see figure~\ref{fig:xi-ry0}b), while the other channels it toward progressively smaller scales in the outer region (see figure~\ref{fig:xi-ry0}c).
As dissipative regions are approached, the flux aligns with the local normal direction and becomes increasingly viscous-dominated. In the limit $r_x, r_z \to 0$, $(\phi_x, \phi_z, \psi) \sim (1,1,0) r$, supplying small scales to offset wall-parallel dissipation, whereas at the wall $(\phi_x, \phi_z, \psi) \sim (0,0,1)Y$, indicating a wall-normal transfer that compensates dissipation associated with wall-normal gradients.

The downward branch becomes predominantly wall-normal for $Y \lesssim h$, revealing a largely scale-independent, top-down transfer ($\psi \gg \phi_x,\phi_z$). Energy produced above the canopy is thus injected into the canopy layer, where it is redistributed and ultimately dissipated in the absence of local production. The scales involved ($r_x \lesssim 0.5$, $r_x^+ \lesssim 471$) are consistent with the KH- and hairpin-type motions identified earlier (\S\ref{sec:pstrain}), indicating that these structures mediate energy transfer between the outer and inner regions.
In contrast, the outer branch remains primarily within the overlying flow and follows a reverse-to-direct cascade sequence: energy first shifts toward larger streamwise separations before bending toward the smallest scales, where it feeds dissipation away from the canopy interface.
Together, these two pathways delineate a clear energetic partition: a top-down transfer that sustains inner-layer fluctuations, and an outer cascade that redistributes production toward small-scale dissipation aloft.

%
Energy transfers in scale space can be separated into viscous and turbulent contributions, highlighting the mechanisms governing the cascade; see figures \ref{fig:xi-ry0}(d-g). The viscous flux acts predominantly at the smallest scales, driving a direct cascade along both $r_x$ and $r_z$ independently of wall-normal position. Near the origin ($r_x, r_z \to 0$), where velocity gradients are strongest, viscous effects dominate and fluxes converge, supplying energy to balance local dissipation.
The turbulent flux is dominant across intermediate and larger scales, representing inertia-driven transfer whose magnitude varies with both scale and distance from the wall. Within the canopy (panels $d$ and $e$), motions drive an inverse cascade toward larger separations in both directions for $0 < r_x, r_z \lesssim 0.5$. Just above the canopy (panels $f$ and $g$), the process becomes scale-dependent: for $r_z \lesssim 0.3$ ($r^{+}_z < 283$), energy first shifts to larger streamwise separations (inverse) before returning to smaller scales (direct), whereas for $r_z \gtrsim 0.4$ ($r^{+}_z > 377$) the inverse transfer dominates both directions. Overall, this decomposition shows that viscous fluxes dominate at the smallest scales, while turbulent fluxes control the energy transfer across intermediate and larger scales through combined direct and inverse cascades.

%
%

\begin{figure}
  \centering
  \includegraphics[width=\textwidth,trim={0cm 0cm 0cm 0cm},clip]{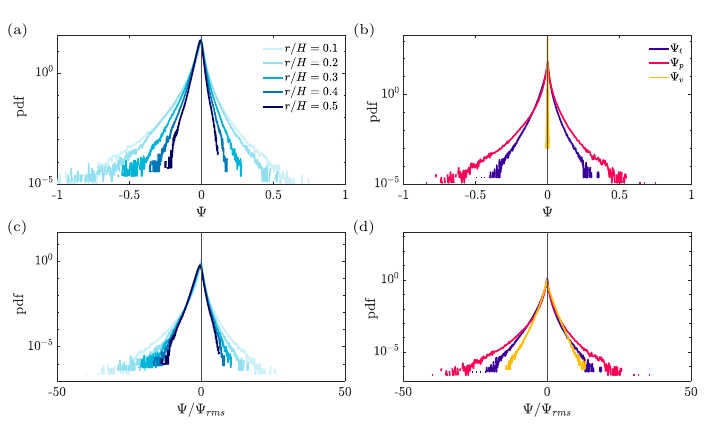} 
  \caption{Distribution of the average physical-space flux $\Psi$ at $Y = 0.26H$ ($Y^{+} = 245$). The flux is obtained by averaging the instantaneous spatial flux over circular areas of radius $r$ in the $r_x$–$r_z$ plane, for different values of $r$.
Panels (a,c): spatial distribution of the total flux $\Psi$ for the scales $r/H = 0.1, 0.2, 0.3, 0.4, 0.5$ ($r^{+} \approx 94, 189, 283, 377, 471$).
Panels (b,d): decomposition of $\Psi$ into its constituent components—the turbulent flux $\Psi_t$, pressure flux $\Psi_p$, and viscous flux $\Psi_v$.}
  \label{fig:psi_pdf}
\end{figure}
We conclude this section by examining the energetic coupling between the outer flow and the canopy via the average downward-directed flux lines (group I in figure \ref{fig:xi-ry0}). To quantify this exchange locally and across scales, we compute the flux at different streamwise $X$ and spanwise $Z$ positions through a circular patch of radius $r$ in the $(r_x,r_z)$ plane at the interface $Y = 0.26 \approx h$:
\begin{equation}
\Psi(X,Y,Z,r) = \frac{1}{S(r)} \oint_{S(r)} \left( v^* \delta q^2 + \frac{2}{\rho_f} \delta p \delta v - \frac{\nu}{2} \frac{\partial}{\partial Y} \delta q^2 \right) \mathrm{d}\Sigma, \quad S(r) = \pi r^2,
\end{equation}
where the three terms correspond to turbulent, pressure, and viscous contributions. Averaging along the circumference yields similar results. Several scales are considered.
The distributions of $\Psi$ (see figure \ref{fig:psi_pdf}) indicate, at all scales, a predominant downward transfer of scale energy from the outer flow into the canopy, sustaining motions within it. Upward transfers also occur locally but less frequently. For instance, at $r = 0.1$, strong downward events ($\Psi < -\Psi_\mathrm{rms}$) occur with probability 0.111, while upward events ($\Psi > \Psi_\mathrm{rms}$) occur with probability 0.048. The intensity and intermittency of these transfers are largest at smaller scales, reflecting the contribution of intense sweep and ejection events associated with hairpin vortex legs ($r \approx 0.25$ or $r^+ \approx 236$), whereas larger KH-like structures ($r \approx 0.5$ or $r^+ \approx 471$) provide a steadier, more continuous energy supply.  
Decomposition into turbulent, pressure, and viscous components confirms that both downward and upward transfers are dominated by turbulent and pressure-mediated mechanisms, with viscous effects playing a minor role at this wall-normal location. This analysis demonstrates that vertical energy exchange between outer and canopy regions is concentrated in intense, intermittent events rather than uniform transport.

\subsubsection{The role of pressure-strain in shaping the canopy-layer fluctuation structure}
\label{sec:pstrain}

\begin{figure}
  \centering
  \includegraphics[width=\textwidth,trim={0cm 0cm 0cm 0cm},clip]{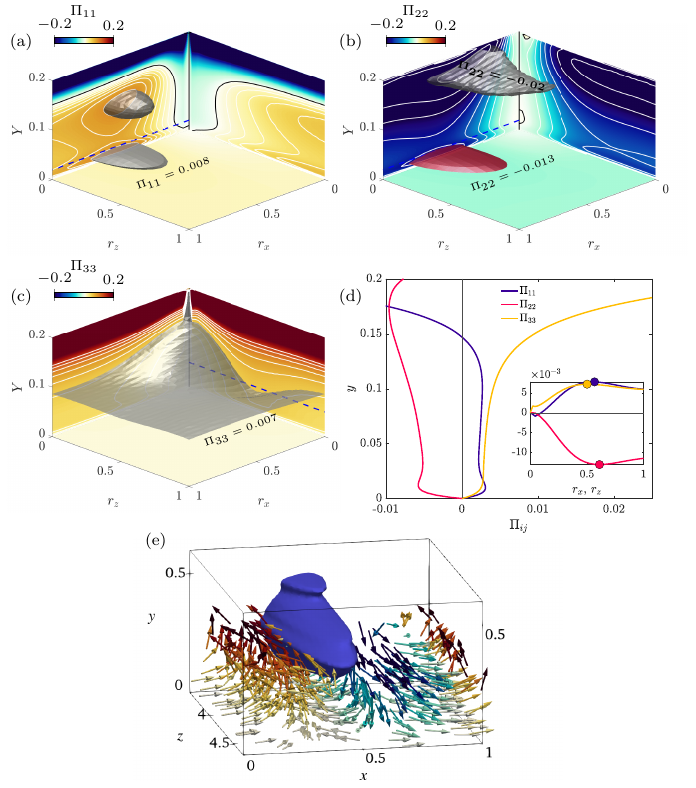}
  \caption{Panels (a-c): pressure-strain term $\Pi_{ij}$ in the $r_y = 0$ plane, zoomed inside the canopy layer for $Y < 0.2$. White lines indicate contours of $\Pi_{ij}$ at various levels, with the black line marking the $\Pi_{ij} = 0$ contour, and are used to identify the local minima/maxima.  
Panel (d) shows the single-point pressure-strain terms as a function of the wall-normal coordinate $Y$.  
Panel (e) illustrates an example of energy redistribution within the canopy layer driven by pressure-strain. The iso-surface corresponds to negative Fourier-filtered pressure fluctuations at $\tilde{p} = -0.35$, superimposed with vectors of the Fourier-filtered velocity field $\boldsymbol{\tilde{u}}$, coloured by the intensity of the vertical component $\tilde{v}$.
}
  \label{fig:pressure-strain-zoom}
\end{figure}

We now examine the mechanism shaping the structural organisation of velocity fluctuations within the canopy layer (\S\ref{sec:structure}). In this region, $\xi < 0$, indicating that the local flow is inactive in producing turbulent energy. Instead, the scale-dependent structure of the fluctuations is inherited from motions above the canopy, primarily through sweep events, and redistributed locally by the pressure--strain term, $\Pi_{ij}$.

Figure~\ref{fig:pressure-strain-zoom} shows the scale-space distribution of $\Pi_{ij}$ within the canopy ($Y \le 0.2$). Across all scales and positions, $\Pi_{22} < 0$ while $\Pi_{11} > 0$ and $\Pi_{33} > 0$. The no-penetration boundary condition enforces an effectively two-dimensional near-wall flow, so vertical velocity fluctuations are reoriented toward wall-parallel directions. Compared to canonical planar channel flows \citep{gatti-etal-2020a}, the region of $\Pi_{22} < 0$ extends further from the wall, reflecting enhanced suppression of vertical motions by the canopy filaments.
Within the canopy, $\Pi_{22}$ exhibits two localised minima at $(Y, r_x, r_z) \approx (0.013, 0, 0.65-0.8)$ and $(0.013, 0.5-0.65, 0)$, coinciding with positive peaks in $\Pi_{11}$ and $\Pi_{33}$ at $(Y, r_x, r_z) \approx (0.013-0.025, 0, 0.4-0.65)$ and $(0.012, 0.45-0.55, 0)$. These locations correspond to the dominant flow scales identified near the canopy tip, see \S\ref{sec:structure} and figure~\ref{fig:dui-duj-RY0-zoom}.  

Despite attenuation by the filaments, the canopy-layer fluctuations retain a scale-dependent structure dictated by the outer flow and mediated by pressure--strain redistribution. Streamwise fluctuations ($u$) reflect the characteristic length scales of KH rolls, while spanwise fluctuations ($w$) preserve the spanwise scales associated with the legs of hairpin-like vortices. This provides a direct mechanistic link between outer-layer dynamics and the observed organisation within the canopy, confirming that pressure--strain-mediated redistribution governs the scale-specific imprint of the outer flow.
An instantaneous snapshot (figure~\ref{fig:pressure-strain-zoom}, bottom panel) illustrates the mechanism: large-scale sweeps generated by KH rolls inject high-momentum fluid into the canopy. Interaction with the filaments and wall redirects these vertical motions via pressure--strain, enhancing wall-parallel $u$-fluctuations at scales consistent with the KH rolls. A similar pattern occurs for $w$-fluctuations at $r_z \ne 0$, highlighting the role of hairpin legs in mediating spanwise momentum redistribution.

\subsection{Flexibility effects} 
\label{sec:flex}

\begin{figure}
  \centering
  \includegraphics[width=\textwidth,trim={0cm 0cm 0cm 0cm},clip]{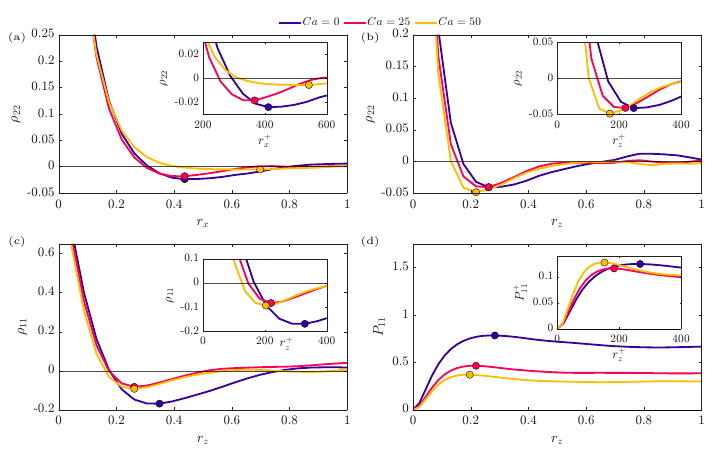}
  \caption{
  Influence of filament flexibility on outer-layer flow scales. Results are shown for three Cauchy numbers: $Ca = 0$, $Ca = 25$, and $Ca = 50$.  
(a) Correlation coefficient $\rho_{22}$ as a function of $r_x$ in the $r_y = r_z = 0$ plane, evaluated at the wall-normal location where $\langle \delta v \delta v \rangle$ peaks.  
(b) Same as (a), but in the $r_x = r_y = 0$ plane.  
(c) Same as (b), but for $\rho_{11}$ and $\langle \delta u \, \delta u \rangle$.  
(d) Turbulent production term $P_{11}$ as a function of $r_z$ in the $r_x = r_y = 0$ plane, evaluated at the wall-normal location corresponding to the peak of $\langle \delta u \, \delta u \rangle$.  
In all panels, insets show the corresponding quantities in viscous units.
}
  \label{fig:ca_outer}
\end{figure}

\begin{figure}
  \centering
  \includegraphics[width=\textwidth,trim={0cm 0cm 0cm 0cm},clip]{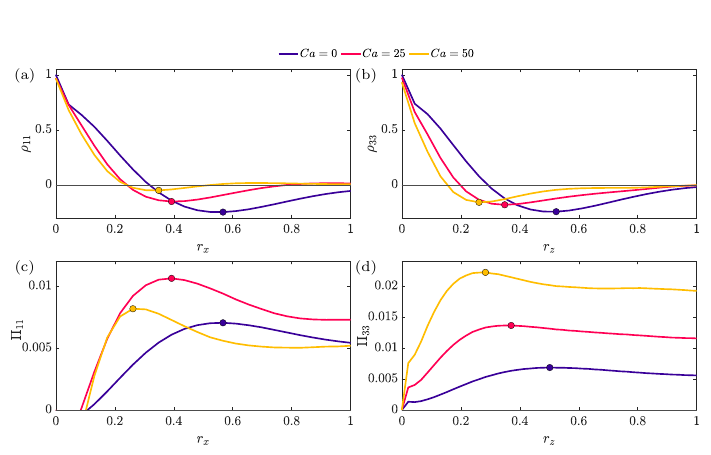}
  \caption{
  Influence of filament flexibility on canopy-layer flow scales. Results are shown for three Cauchy numbers: $Ca = 0$, $Ca = 25$, and $Ca = 50$.  
(a) Correlation coefficient $\rho_{11}$ as a function of $r_x$ in the $r_y = r_z = 0$ plane, evaluated at $Y = 0.03H$. 
(b) Same as (a), but for $\rho_{33}$ in the $r_x = r_y = 0$ plane.  
(c) Pressure-strain term $\Pi_{11}$ as a function of $r_x$ in the $r_x = r_y = 0$ plane, evaluated at $Y = 0.03H$.  
(d) Same as (c), but for $\Pi_{33}$ in the $r_y = r_z = 0$ plane.
}
  \label{fig:ca_inner}
\end{figure}

We now assess the role of filament flexibility across $Ca \in [0,50]$. While the dominant physical mechanisms and key flow structures remain broadly consistent across canopy configurations, increasing $Ca$ introduces systematic quantitative differences. As shown in figure~\ref{fig:first_order_stats}(b), higher $Ca$ enhances turbulent kinetic energy within the canopy due to increased filament compliance, while energy above the canopy is slightly reduced, reflecting a weaker and more diffuse shear layer.  

To quantify these effects, we focus on two aspects. First, we assess the coherence of the flow structures as $Ca$ varies, revealing that more flexible canopies tend to partially decorrelate the KH-like rolls and hairpin structures, particularly near the canopy tips. Second, we examine how the canopy modulates vertical energy transfer from the outer flow into the interior. Increased filament flexibility effectively creates a partially shielding layer, attenuating the penetration of outer-layer motions and modifying the scale-dependent energy fluxes within the canopy. This analysis highlights that while the flow retains its qualitative organisation, the quantitative redistribution of turbulent energy across scales is sensitive to canopy compliance, with implications for both momentum transport and the imprint of coherent structures inside the canopy.


The influence of $Ca$ on outer-layer flow structures is quantified in figure~\ref{fig:ca_outer}, which reports the two-point correlation coefficients $\rho_{11}$ and $\rho_{22}$ as functions of $r_x$ and $r_z$, evaluated at the wall-normal locations corresponding to the peaks of the respective structure functions. With increasing $Ca$, the peak of $\rho_{22}(r_x)$ (figure~\ref{fig:ca_outer}a), associated with the KH rolls, decreases in magnitude and shifts toward larger $r_x$ and lower wall-normal positions. This reflects a weakening and broadening of the shear layer, producing KH rolls that are less coherent and more diffuse. For $Ca = 50$, $\rho_{22}(r_x)$ becomes only marginally negative, indicating near-complete suppression of KH activity.  
Figures~\ref{fig:ca_outer}(b–c) show $\rho_{22}$ and $\rho_{11}$ as functions of $r_z$, capturing the statistical footprints of streamwise-oriented vortices and the low- and high-speed streaks (LSS/HSS). The coherence and intensity of the LSS/HSS decrease with increasing $Ca$ \citep[consistent with][]{sundin-bagheri-2019}, with a slight reduction of the spanwise scale in both outer and viscous units. In contrast, the signature of streamwise-oriented vortices is only marginally affected by canopy flexibility, with the peak of $\rho_{22}$ slightly moving towards smaller spanwise scales as $Ca$ increases.  
The attenuation of the LSS/HSS is linked to a reduction in turbulent production $P_{11}$, resulting from the weakened shear layer (i.e., smaller $\mathrm{d}U/\mathrm{d}y$, see figure~\ref{fig:first_order_stats}a). In the most flexible case ($Ca = 50$), the secondary peak in $\langle \delta u \delta u \rangle$ in the $r_x = r_y = 0$ plane, prominent in the rigid canopy (figure~\ref{fig:dudu-sec-peak}), disappears, further confirming the near-complete suppression of KH rolls. 

Within the canopy layer, the flow remains dominated by streamwise and spanwise fluctuations for all $Ca$ considered, whose organisation reflects the imprint of the overlying motions near the canopy tips (see figure~\ref{fig:pressure-strain-zoom}). Figures~\ref{fig:ca_inner}(a,b) show $\rho_{11}$ and $\rho_{33}$ as functions of $r_x$ and $r_z$ at $Y = 0.03$, with similar trends for small variations in $Y$. Fluctuation intensity increases with $Ca$ across all scales, accompanied by stronger pressure-strain redistribution (figure~\ref{fig:first_order_stats}, $\Pi_{11}$ and $\Pi_{33}$ in figures~\ref{fig:ca_inner}c,d), indicating enhanced reorientation of wall-normal motions into wall-parallel components, likely due to stronger vertical confinement. At the same time, the spatial coherence of the fluctuations decreases with flexibility, reflecting a loss of structural organisation consistent with the attenuation of outer-layer motions.

\begin{figure}
  \centering
  \includegraphics[width=\textwidth,trim={0cm 0cm 0cm 0cm},clip]{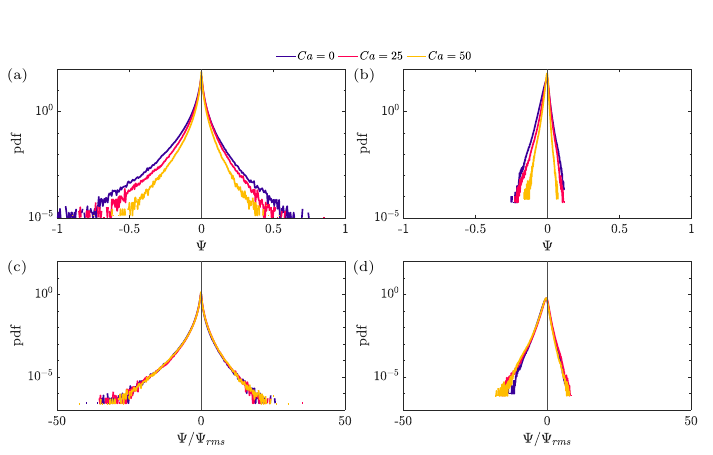}
  \caption{Influence of filament flexibility on the distribution of the average physical-space flux $\Psi$, evaluated at the mean canopy tip location $Y = \langle y_{\mathrm{tip}} \rangle_f$ for each case. The flux is computed as described in \S\ref{sec:agke-results}. Results are shown for three Cauchy numbers: $Ca = 0$, $Ca = 25$, and $Ca = 50$.  
(a, c) Spatial distribution of the total flux $\Psi$ at scale $r/H = 0.1$.  
(b, d) Spatial distribution of the total flux $\Psi$ at scale $r/H = 0.5$.
}
  \label{fig:flux_vs_Ca}
\end{figure}

We conclude this section by examining how filament flexibility affects the energetic coupling between the outer flow and the canopy. Figure~\ref{fig:flux_vs_Ca} reports the scale-averaged flux $\Psi$ at two representative scales, $r = 0.1$ and $r = 0.5$, evaluated at the mean filament-tip height. For all $Ca$, the distributions remain left-skewed, indicating a net top-to-bottom energy transfer across scales (see \S\ref{sec:fluxes}). 
Increasing $Ca$ reduces the magnitude of the distribution tails (figures~\ref{fig:flux_vs_Ca}a–b), reflecting a weaker shear layer and a loss of coherence in sweep–ejection structures. Energy transfer is strongest at smaller scales, where the influence of the flexibility is most pronounced. When normalised by the root-mean-square value (figures~\ref{fig:flux_vs_Ca}c–d), the distributions collapse across all $Ca$, indicating that the relative intermittency of energy transfer remains largely unaffected by filament flexibility.

\section{Conclusions}
\label{sec:conclusions}

\begin{figure}
  \centering
  \includegraphics[width=0.9\textwidth]{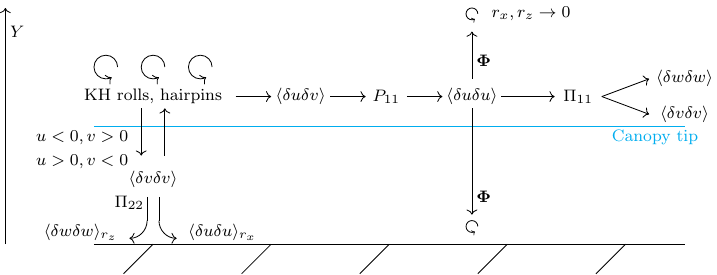}
  \caption{Schematic representation of the mechanism sustaining velocity fluctuations in canopy turbulence.}
  \label{fig:energy-scheme}
\end{figure}

This study has addressed the central question of how turbulence is organised and sustained in dense, submerged canopy flows, and how energy is exchanged across scales and between the canopy layer and the overlying shear flow. Using the DNS database of \citet{monti-olivieri-rosti-2023} and the anisotropic generalised Kolmogorov equations (AGKE) \citep{gatti-etal-2020a}, we have provided a space- and scale-resolved characterisation of production, redistribution, and inter-scale transfer for both rigid and flexible canopies at $Re_b = 5000$ and Cauchy numbers $Ca = 0$, $25$, and $50$.

Figure \ref{fig:energy-scheme} summarises the main energy pathways across scales and flow regions. Energy production is localised at the interfacial shear layer, over a narrow range of streamwise and spanwise scales aligned with the dominant coherent structures in the outer layer, such as Kelvin--Helmholtz-like rolls, hairpin vortices and streaks. Within the canopy, fluctuations are sustained non-locally, primarily through inter-scale transfer and pressure--strain redistribution, inheriting their organisation and characteristic scales from the outer-layer motions. Sweep events induced by the outer-layer structures play a key role in transferring energy downward, while pressure--strain interactions redistribute it among velocity components, shaping the anisotropy of canopy-layer fluctuations.

The energy exchange across the canopy interface is asymmetric but bidirectional. While the net transfer is from the outer layer into the canopy, intermittent reverse fluxes occur at all scales, with the strongest events dominated by small-scale motions. This interplay establishes a dynamic coupling between inner and outer regions, linking coherent outer-layer structures to the organisation of canopy turbulence. Vertical energy exchange at the canopy interface is therefore concentrated in intense, intermittent events rather than uniform transport. In contrast, velocity fluctuations in the bulk flow are sustained by upward transfers originating from the shear layer above the canopy tip, combining both inverse and direct cascades toward the smallest scales.

Canopy flexibility modifies both outer- and inner-layer dynamics. Increasing compliance weakens the coherence of outer-layer structures, diffuses the shear layer, and reduces the efficiency of inter-layer energy exchanges. Within the canopy, although the overall energy increases, the coherence of structures inherited from the motions above is diminished, and the scale-dependent coupling with the outer flow is attenuated, leading to weaker energy exchanges across the canopy interface.

Overall, this study provides a unified energetic interpretation that connects coherent structures to production, redistribution, and inter-scale transfer mechanisms. Future work could extend this analysis to broader parameter ranges, for example by exploring sparse canopies, where large-scale structures penetrate deeper into the canopy layer \citep{sharma-garcia-mayoral-2020}, varying canopy height, or artificially constraining filaments to be rigid to isolate the effect of reconfiguration \citep{foggirota-etal-2024}. It would also be of interest to investigate non-uniform or patchy canopy arrangements \citep{lohrer-frohlich-2023, foggirota-etal-2025}, where added inhomogeneity is expected to significantly alter the scale-resolved energy transfer mechanisms identified in this study.

\backsection[Acknowledgements]{
The research was supported by the Okinawa Institute of Science and Technology Graduate University (OIST) with subsidy funding to M.E.R. from the Cabinet Office, Government of Japan. M.E.R.~also acknowledges funding from the Japan Society for the Promotion of Science (JSPS), grants 24K00810 and 24K17210. R.B. acknowledges partial funding from Politecnico di Milano. The authors acknowledge the computer time provided by the Scientific Computing and Data Analysis section of the Core Facilities at OIST and the computational resources offered by the HPCI System Research Project with grants hp210025, hp220402, hp240006, hp250035.
}

\backsection[Declaration of interests]{The authors report no conflict of interest.}

\backsection[Author ORCIDs]{\\
Riccardo Bertoncello \orcidA{}: \url{https://orcid.org/0009-0000-7023-1267} \\
Alessandro Chiarini \orcidB{}: \url{https://orcid.org/0000-0001-7746-2850} \\
Giulio Foggi Rota \orcidC{}: \url{https://orcid.org/0000-0002-4361-6521}; \\
Maurizio Quadrio \orcidD{}: \url{https://orcid.org/0000-0002-7662-3576}; \\
Marco Edoardo Rosti \orcidE{}: \url{https://orcid.org/0000-0002-9004-2292}; \\
}

\appendix
\section{Fluid velocity on the filaments}
\label{appA}

\begin{figure}
   \centering
   \includegraphics[width=.57\textwidth]{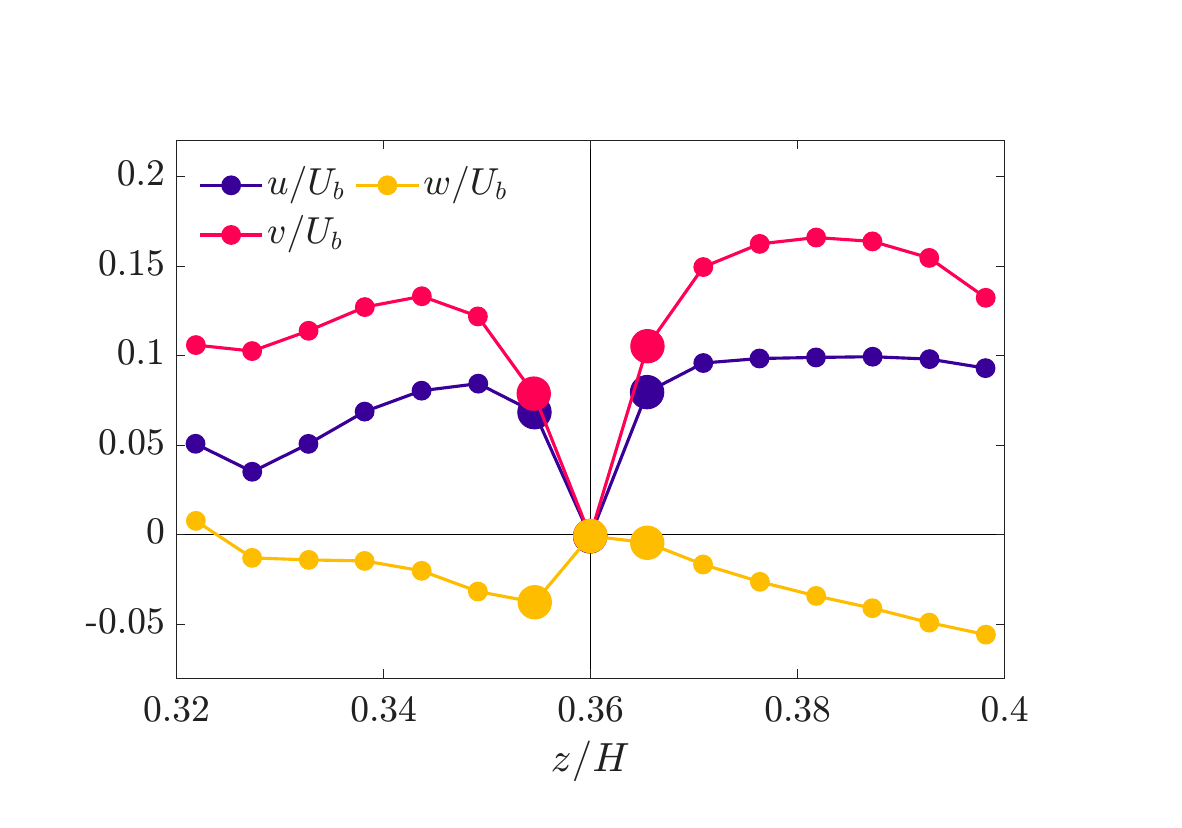}
   \caption{Instantaneous velocity components extracted along a spanwise line passing through the representative rigid stem located at $(x/H, z/H) = (0.11, 0.36)$. All three velocity components vanish exactly at the filament coordinate. Large markers indicate the Eulerian grid points within the compact support of the IBM forcing.}
   \label{fig:noVel}
\end{figure}

In this appendix, we show that the immersed boundary method employed in the present study \citep{yu-2005, huang-shin-sung-2007} rigorously enforces the no-slip and no-penetration boundary conditions at the Lagrangian points representing the canopy filaments. To provide the most stringent test, we consider the configuration featuring rigid filaments, as this induces the strongest hydrodynamic back-reaction on the fluid. We extract the velocity field from an instantaneous snapshot along a spanwise line (parallel to the $z$-axis) passing directly through a representative stem located at $(x/H, z/H) = (0.11, 0.36)$. As shown in figure \ref{fig:noVel}, all three velocity components ($u, v, w$) vanish exactly at the coordinate of the Lagrangian point, confirming that the required kinematic constraints are strictly satisfied. The Lagrangian force responsible for enforcing this condition is projected onto the underlying Eulerian fluid grid over a three-point compact support via convolution with a regularised delta function \citep{roma-peskin-berger-1999}.

\bibliographystyle{jfm}

\end{document}